\documentclass[doublecol]{epl2}
\title{Improving information filtering via network manipulation}
\usepackage{amssymb}
\usepackage{xcolor}
\author{Fuguo Zhang$^{1,2}$ and An Zeng$^{2}$\footnote{an.zeng@unifr.ch}}
\shortauthor{Fuguo Zhang and An Zeng}

\institute{$^{1}$ School of Information and Technology, Jiangxi University of Finance and Economics, Nanchang 330013, P.R. China\\
$^{2}$ Department of Physics, University of Fribourg, Chemin du Mus\'{e}e 3, CH-1700 Fribourg, Switzerland\\
}

\abstract{Recommender system is a very promising way to address the problem of overabundant information for online users. Though the information filtering for the online commercial systems received much attention recently, almost all of the previous works are dedicated to design new algorithms and consider the user-item bipartite networks as given and constant information. However, many problems for recommender systems such as the cold-start problem (i.e. low recommendation accuracy for the small degree items) are actually due to the limitation of the underlying user-item bipartite networks. In this letter, we propose a strategy to enhance the performance of the already existing recommendation algorithms by directly manipulating the user-item bipartite networks, namely adding some virtual connections to the networks. Numerical analyses on two benchmark data sets, \emph{MovieLens} and \emph{Netflix}, show that our method can remarkably improve the recommendation performance. Specifically, it not only improve the recommendations accuracy (especially for the small degree items), but also help the recommender systems generate more diverse and novel recommendations.}

\begin{document}

\maketitle

\section{Introduction}
In the Internet era, the rapid growth of the World-Wide-Web leads to a serious problem of information overload: people are now facing too many choices to be able to find out those most relevant ones~\cite{Broder2000}. So far, the most promising way to efficiently filter the abundant information is to employ the personalized recommendations~\cite{Adomavicius05,Cacheda11}. That is to say, using the personal history record
of a user to uncover his preference and to return each user with the most relevant items according to his taste~\cite{PRlinyuan}. For instances, youtube.com uses people's video viewing record to provide individual suggestions for their potential interested videos.

There are already many recommendation algorithms for the online user-item commercial systems. Among these algorithms, the simplest one is the popularity-based recommendations, which recommend the most popular items to users. However, such recommendations are not personalized so that identical items are recommended to individuals with far different tastes. By comparison, the collaborative filtering makes use of collective data from individual preferences
to provide personalized recommendations\cite{Konstan1997,Herlocker04}. Recently, recommendation algorithms have been proposed from a
physics perspective. For example, the process of
mass diffusion (MD) was applied on the user-item bipartite networks to explore items of potential interest for a user~\cite{Zhou07}.
The mass diffusion  algorithm outperforms the previous ones in the recommendation accuracy.
However, such method is still biased to popular items even if individual preferences are considered. An alternative approach,
based on the heat conduction (HC) on the user-item graphs,
was thus introduced~\cite{Zhang07}. This algorithm provides users with many novel items and leads to
diverse recommendations among users.
However, HC has low accuracy compared with MD.
This drawback is eventually solved by combining MD with HC in a hybrid approach,
which can be well-tuned to obtain significant improvement in both recommendation accuracy and
item diversity~\cite{Zhou10}. More Recently, the long term influence of such hybrid approach on network evolution has been studied~\cite{EPL9718005}.

However, all these methods are focusing on improving the recommendation all from the system point of view. The recommendation on the items with little information are actually still a critical challenging~\cite{PRlinyuan}. For the fresh or unpopular items (also called niche items), it is very difficult to predict the potential users who are going to interest in them due to lacking of historical record. Such problem is always referred as item cold-start problem and many researches have been dedicated to solve this problem. Related works concerning this issue are mainly based on modifying the existing methods by introducing some parameters~\cite{PRE83066119,EPL9558003,PRE84037101}.

Different from previous works, we tried to solve the item cold-start problem through a very fundamental way in this letter. Instead of designing a new recommendation algorithm, we solve the item cold-start problem by directly manipulating the underlying user-item bipartite networks~\cite{Evans07,PhysicaA3911822}. Actually, the idea of the network manipulation has been applied to enhance many kinds of network functions such as synchronization~\cite{PNAS10710342,NJP17083006}, traffic dynamics~\cite{EPL8958002}, percolation~\cite{PNAS1083838,PRE85066130}, navigation~\cite{PRL104018701} and so on. In our case, we first analyze the historical record of each item and accordingly add some virtual connections to the networks (especially for the small degree items) to provide the recommendation algorithm with more information. By using the MD and the hybrid algorithms, we find that the recommendation accuracy for the small degree items can be largely enhanced after manipulating the networks. The further test on the overall recommendation metrics, our method are shown to help the MD and the hybrid algorithms to significantly improve the recommendation diversity.

\section{Recommendation algorithms}

Online commercial systems can be well described by the user-item bipartite networks. If a user collects an item, a link is drawn between them. Specifically,
we consider a system of $N$ users and $M$ items represented by a bipartite network with adjacency matrix $A$, where the element $a_{i\alpha}=1$ if a user $i$ has collected an item $\alpha$, and $a_{i\alpha}=0$ otherwise (throughout this paper we
use Greek and Latin letters, respectively, for item- and user-related indices).

There are many recommendation algorithms. In this letter, we mainly consider the Mass Diffusion (MD), Heat Conduction (HC) and the corresponding hybrid algorithms of these two algorithms (Hybrid). We first briefly describe these algorithms.

For a target user $i$, the MD algorithm~\cite{Zhou07} starts by
assigning one unit of resource to each item collected by $i$, and redistributes the resource through the user-item network. We denote the vector $\textbf{f}^{i}$ as the initial resources on items, where the $\alpha$-th component $f_{\alpha}^{i}$ is the resource possessed by item $\alpha$. Recommendations for the user $i$ are obtained
by setting the elements in $\textbf{f}^{i}$ to be $f^{i}_{\alpha}=a_{i\alpha}$, in accordance with the items the user has
already collected.
The redistribution is represented by $\widetilde{\textbf{f}^{i}}=W\textbf{f}^{i}$, where
\begin{equation}
W_{\alpha\beta}=\frac{1}{k_{\beta}}\sum\limits_{j=1}^{N}\frac{a_{j\alpha}a_{j\beta}}{k_{j}},
\end{equation}
is the diffusion matrix, with $k_{\beta}=\sum^{N}_{l=1}a_{l\beta}$ and $k_{j}=\sum^{M}_{\gamma=1}a_{j\gamma}$ denoting the degree of item $\beta$ and user $j$ respectively. The resulting recommendation
list of uncollected items is then sorted according to $\widetilde{f}^{i}_{\alpha}$ in descending order.
Physically, the diffusion is equivalent to a three-step random walk starting with $k_i$ units of resources on the target user $i$ and the process can be seen in fig. 1(a). The \emph{recommendation score} of an item is taken to be its amount of gathered resources after the diffusion. This algorithm was shown to enjoy a high recommendation accuracy.

The HC algorithm~\cite{Zhang07} works similar to the MD algorithm, but instead follows a conductive process represented by
\begin{equation}
W_{\alpha\beta}=\frac{1}{k_{\alpha}}\sum\limits_{j=1}^{N}\frac{a_{j\alpha}a_{j\beta}}{k_{j}}.
\end{equation}
Physically, the recommendation scores can be interpreted as the temperature of an item, which is the average temperature of its nearest neighborhood, i.e. its connected users. The higher the temperature of an item, the higher its recommendation score. The HC process can be seen in fig. 1(b). By using this algorithm, the items with small degree can receive relatively high recommendation score and finally be promoted to appear in the top recommendation list.

The hybrid algorithm of MD and HC was proposed in~\cite{Zhou10}, with the new recommendation score $\widetilde{h}_{\alpha}$ given by
\begin{equation}
\label{eq_hybrid}
\widetilde{h}_{\alpha}=\lambda\frac{\widetilde{f}^{\rm MD}_{\alpha}}{\rm Max(\widetilde{f}^{\rm MD})}+\left(1-\lambda\right)\frac{\widetilde{f}^{\rm HC}_{\alpha}}{ \rm Max(\widetilde{f}^{\rm HC})}.
\end{equation}
where the parameter $\lambda$ adjusts the relative weight between the two algorithms. When $\lambda$ increases from $0$ to $1$, the hybrid algorithm changes gradually from HC to MD. Such hybrid approach was shown to achieve both accurate and diverse recommendation.

\section{Data}
In order to test the performance of the recommendation results, we use two benchmark
data sets in this letter. The first one is the MovieLens data~\cite{movielens} which
has $1,682$ movies (items) and $943$ users. The other is Netflix data~\cite{netflix} consisting of $10,000$ users and $6,000$ movies. The data sets are random samplings of users activity records in these two online systems. In both data sets, users can vote movies by giving different rating levels from 1 to 5 (i.e. worst to best). Here, only the rating larger than 2 are considered as a link. After this preliminary filtering, there are finally $82,520$ links in movielens data and $701,947$ links in the netflix data. Each data is then randomly divided into two parts: the training set ($E^T$) and the probe set ($E^P$). The training set contains $90\%$ of the original data and the recommendation algorithm runs on it. The probe set has the remaining $10\%$ of the data and will be used to test the performance of the recommendation results. In our simulation, we actually try several different divisions of the $E^T$ and $E^P$, and the results are quite robust.

\section{The network manipulating method}
\begin{figure}
  \center
  \includegraphics[width=0.85\columnwidth]{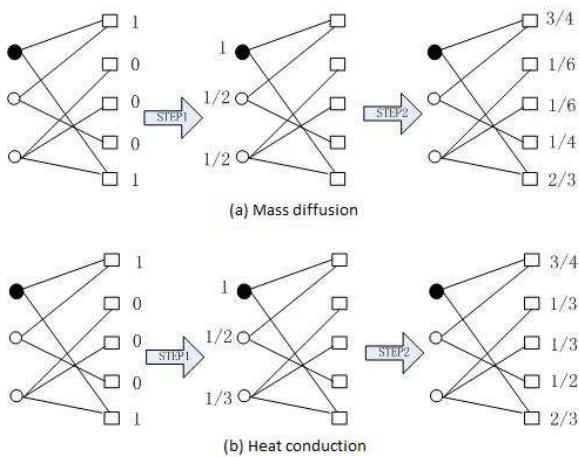}
\caption{(Color online) The illustration of the (a) MD and (b) HC algorithms on the bipartite user-object network. Users are shown as circles; objects are squares. The target user is indicated by the shaded circle.}\label{fig1}
\end{figure}

\begin{figure}
  \center
  \includegraphics[width=\columnwidth]{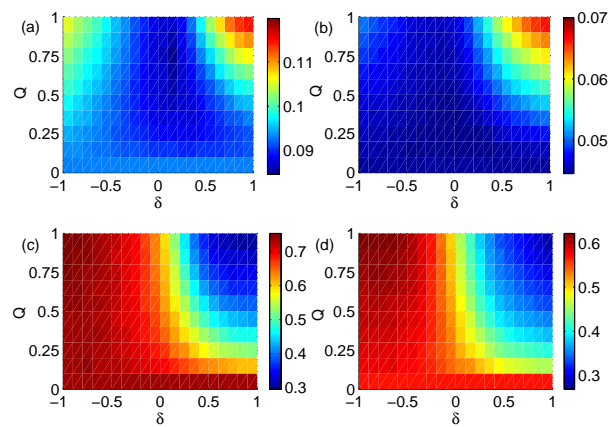}
\caption{(Color online) The overall ranking score $R$ under different $\delta$ and $Q$ in (a) Movielens and (b) Netflix, and the local ranking score $R_{k\leq5}$ under different $\delta$ and $Q$ in (c) Movielens and (d) Netlfix.}\label{fig2}
\end{figure}

The network manipulating (NM) method takes place after dividing the data to $E^P$ and $E^T$. The main idea of NM is to add some virtual links to the training set $E^T$, so that the niche items will have more information to provide to the recommendation algorithm. Denoting $Q$ as the fraction of links added, the total number of virtual links will be $Q|E^T|$. For each item, the probability to receive virtual links is related to its degree, i.e. $p_{\alpha}\propto k_{\alpha}^{-\delta}$ where $\delta$ is a tunable parameter. when $\delta>0$, the items with smaller degree tend to receive more links, and vice versa. Supposing an item $\alpha$ is selected to receive a link, the virtual link will connect to the user who enjoys the highest average similarity to the already existing selectors of the item $\alpha$. In this letter, the similarity is calculated by \emph{Salton Index}~\cite{cosineindex} as
\begin{equation}
s_{ij}=\frac{|\Gamma(i)\cap\Gamma(j)|}{\sqrt{k_{i}k_{j}}}.
\end{equation}
where $\Gamma(i)$ denotes the set of neighbors of user $i$. After adding virtual connections to the networks, we will employ the MD revision and the Hybrid algorithms to do the recommendation and the combinations are denoted as ``MD with NM" and ``Hybrid with NM", respectively.

In the NM method, the added virtual links may happen to be one of links in $E^P$. After running the recommendation process, the Mass diffusion and Hybrid algorithms will generate the final recommendation list by sorting the recommendation score of all the uncollected items for the target user. In our simulation, if there is a virtual link connecting the target user and an item, this item is still considered as an uncollected item for the target user. Therefore, the NM method does not reduce the number of candidate items for recommendation (i.e. the ratio of probe-set links out of all non-existing links stays unchanged).

We remark that the NM method is quite efficient in sparse network. For each item, the NM method has to calculate the average Salton similarity between the users who already selected it and the users who haven't selected it. Since calculating the Salton similarity between users costs $O(\bar{k}_{i})$, the computational complexity for the NM procedure is $O(M\bar{k_{\alpha}}(N-\bar{k_{\alpha}})\bar{k_{i}})$. Since $\bar{k_{\alpha}}<<N$, it can be further written as $O(M\bar{k_{\alpha}}(N-\bar{k_{\alpha}})\bar{k_{i}})\approx O(M\bar{k_{\alpha}}N\bar{k_{i}})=O((M*E*N*E)/(M*N))=O(E^2)$, where $E$ is the total number of links in the network.

\section{Metrics for recommendation}

An effective recommendation should be able to accurately find the items that users like. In order to measure the recommendation accuracy, we make use of
\emph{ranking score} ($R$). Specifically, $R$ measures whether the ordering of the items in the recommendation list matches the users' real preference. As discussed above, the recommender system will provide each user with a ranking list which contains all his uncollected items. For a target user $i$, we calculate the position for each of his link in the probe set. Supposing one of his uncollected item $\alpha$ is ranked at the $5$th place and the total number of his uncollected items is $100$, the ranking score $R_{i\alpha}$ will be $0.05$. In a good recommendation, the items in the probe set should be ranked higher, so that $R$ will be smaller. Therefore, the mean value of the $R$ over all the user-item relations in the probe set can be used to evaluate the recommendation accuracy as
\begin{equation}
R=\frac{1}{|E^{P}|}\sum_{i\alpha\in E^{P}}R_{i\alpha}.
\end{equation}
The smaller the value of $R$, the higher the recommendation accuracy.

\begin{table*}[htbp]
\begin{center}
\caption{The performance of the original MD and the MD with NM methods in \emph{Movielens} and \emph{Netflix} data. The recommendation list length is set as $L=20$. In the MD with NM method, the parameters are $Q=0.8$, $\delta=0.1$ in movielens and $Q=0.2$, $\delta=0.1$ in netflix. The entries corresponding to the best performance over all methods are emphasized in black.}
\label{Table1}
\begin{tabular}{cccc cccc cccc cccc cccc cccc cccc}
\\
\hline
\hline
Network & Method & $R$ &$R_{k\leq5}$ &$P(20)$ &$H(20)$ &$N(20)$\\
\hline
            & Original MD  & 0.0933 & 0.7324 & 0.1427 & 0.7161 & 303.8\\
Movielens   & MD with NM  & \textbf{0.0861} & \textbf{0.5430} & \textbf{0.1451} & \textbf{0.7437} & \textbf{293.3}\\

 \hline
            & Original MD  & 0.0452 & 0.5695 & 0.0808 & 0.5470 & 2841.6\\
Netflix     & MD with NM  & \textbf{0.0446} & \textbf{0.4986} & \textbf{0.0821} & \textbf{0.5618} & \textbf{2829.8}\\
\hline
\hline
\end{tabular}
\vspace*{0.0cm}
\end{center}
\end{table*}

In reality, online systems only present users with only the top part of the recommendation list. Therefore, we further consider another more practical recommendation accuracy measurement called \emph{precision}, which only takes into account each user's top-$L$ items in the recommendation list. For each user $i$, his precision of recommendation is calculated as
\begin{equation}
P_{i}(L)=\frac{d_{i}(L)}{L},
\end{equation}
where $d_{i}(L)$ represents the number of user $i$'s deleted links contained in
the top-$L$ places in the recommendation list. For the whole system, the precision $P(L)$ can be obtained by averaging the individual precisions
over all users with at least one link in the probe set.

\begin{figure}
  \center
  \includegraphics[width=\columnwidth]{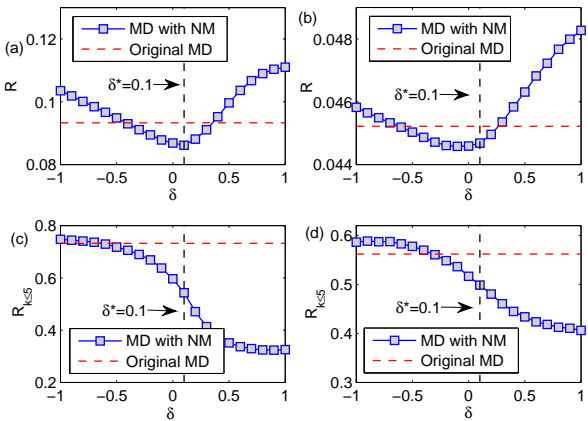}
\caption{(Color online) The overall ranking score $R$ and the local ranking score $R_{k\leq5}$ of the MD with NM method under different $\delta$ when $Q=0.8$ in (a), (c) Movielens and $Q=0.2$ in (b), (d)  Netflix. The vertical dash line is the optimal $\delta$ we used.}\label{fig3}
\end{figure}

\begin{figure}
  \center
  \includegraphics[width=\columnwidth]{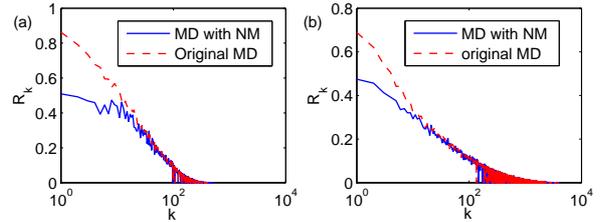}
\caption{(Color online) $R_{k}$ vs. item degree $k$ when using different methods in (a) Movielens and (b) Netflix data sets. The parameters for the MD with NM method are $Q=0.8$, $\delta=0.1$ in movielens and $Q=0.2$, $\delta=0.1$ in netflix.}\label{fig4}
\end{figure}

Predicting what a user likes from the list of best sellers is generally easy in recommendation, while uncovering users' very personalized preference (i.e. uncovering the unpopular items in the probe set) is much more difficult and important. Therefore, diversity should be considered as another significant aspects for recommender systems besides accuracy. In this letter, we employ two kinds of diversity measurement: \emph{interdiversity} and \emph{novelty}.

The interdiversity mainly consider how users' recommendation lists are different from each other. Here, we measure it by the Hamming distance. Denoting $C_{ij}(L)$ as the number of common items in the top-$L$ place of the recommendation list of user $i$ and $j$, their hamming distance can be calculated as
\begin{equation}
H_{ij}(L)=1-\frac{C_{ij}(L)}{L}.
\end{equation}
Clearly, $H_{ij}(L)$ is between $0$ and $1$, which are respectively corresponding to the cases where $i$ and $j$ having the same or entirely different recommendation lists. Again, averaging $H_{ij}(L)$ over all pairs of users, we obtain the mean hamming distance $H(L)$. A more personalized recommendation results in a higher $H(L)$.

The novelty measures the average degree of the items in the recommendation list. For those popular items, users may already get them from other channels. However, it's hard for the users to find the relevant but unpopular item. Therefore, a good recommender system should prefer to recommend small degree items. The metric \emph{novelty} can be expressed as
\begin{equation}
N_{i}(L)=\frac{1}{L}\sum_{\alpha\in O^{i}}k_{\alpha}
\end{equation}
where $O^{i}$ represents the recommendation list for user $i$. A low mean popularity $N(L)$ for the whole system indicates a high novel and unexpected recommendation of items.

\section{Results}

\begin{table*}[htbp]
\begin{center}
\caption{The performance of the Original Hybrid and the Hybrid with NM methods in \emph{Movielens} and \emph{Netflix} data. The recommendation list length is set as $L=20$. In the Hybrid with NM method, the parameters are $Q=0.1$, $\delta=0.1$ in movielens and $Q=0.1$, $\delta=0.1$ in netflix. The entries corresponding to the best performance over all methods are emphasized in black.}
\label{Table1}
\begin{tabular}{cccc cccc cccc cccc cccc cccc cccc}
\\
\hline
\hline
Network & Method & $R$ &$R_{k\leq5}$ &$P(20)$ &$H(20)$ &$N(20)$\\
\hline
            & Original Hybrid  & 0.0759 & 0.5059 & 0.1532 & 0.8055 & 276.7\\
Movielens   & Hybrid with NM  & \textbf{0.0755} & \textbf{0.4898} & \textbf{0.1549} & \textbf{0.8173} & \textbf{269.9}\\

 \hline
            & Original Hybrid  & \textbf{0.0446} & 0.5532 & 0.0810 & 0.5491 & 2839.1\\
Netflix     & Hybrid with NM  & 0.0447 & \textbf{0.5232} & \textbf{0.0816} & \textbf{0.5563} & \textbf{2834.7}\\

\hline
\hline
\end{tabular}
\vspace*{0.0cm}
\end{center}
\end{table*}

We will begin our analysis with comparing the ``MD with NM" and the original MD algorithm. We first investigate the result of ranking score $R$ under different $\delta$ and $Q$. Since the NM method partially aims at solving the item cold-start problem, we also define a local ranking score which is the average ranking score of the items with degree not larger than $5$ (denoted as $R_{k\leq5}$). The results on Movielens and Netflix data are reported in Fig. 2. Clearly, with more links added to the network, $R_{k\leq5}$ becomes smaller. Consequently, the overall $R$ is improved. Given the value of $Q$, a smaller $\delta$ yields a lower $R_{k\leq5}$, which means the recommendation for the small degree items becomes more accurate. However, the overall $R$ does not monotonously change with $\delta$. For each $Q$, there is a corresponding optimal $\delta$ which yields the best $R$.

For solving the item cold-start problem, fig. 2(c) and (d) suggest that large $Q$ generally works better. However, in order to keep a reasonable computational time of the recommendation algorithm, we select $Q=0.8$ in movielens data and $Q=0.2$ in Netflix data. The results of the ranking score are studied more detailedly in fig. 3. In this figure, besides the curve of the MD with NM, we plot the results of the original MD (without adding any virtual links) as a comparison. For overall $R$, $\delta$ near $0$ clearly works better. However, positive $\delta$ is beneficial for improving the ranking score for small degree items. In our simulation, we find $\delta^*=0.1$ is the best trade-off between $R$ and $R_{k\leq5}$. As we can see from fig. 3, under this $\delta^*$, the overall $R$ can outperform the original MD algorithm in both data sets. Moreover, the MD with MN method enjoys a lower $R_{k\leq5}$ than the original MD algorithm in both data sets. For the value of each metric, see table I.

To show how the ranking score varies on items with different value of degrees, we additionally investigate an item-degree-dependent ranking score $R_{k}$~\cite{EPL8158004}. $R_{k}$ is defined as the average ranking score over items with the same value of degrees. In fig. 4, the relation between $R_{k}$ and the item degree $k$ is displayed respectively for the Movielens and Netflix at the optimal parameters $\delta^*=0.1$. Besides the MD with NM method, we also plot the results of original MD for comparison. Obviously, the ranking score of small degree items can be significantly improved by adding the virtual connections. Moreover, the ranking score of large degree items can be effectively preserved.

As discuss above, another way to estimate the accuracy of the recommendation results is the precision. We select the recommendation list $L=20$, and report the precision of MD with NM, original MD algorithm in table I. In additional to accuracy, the recommendation diversity is of great significance. For interdiversity, we can estimate how the recommendation results are different from user to user. A larger hamming distance indicates a more personalized recommendation. Besides, the novelty is also an important aspect. With a small novelty, the average degree of the recommended items are low, so that more fresh items will appear in the recommendation list. Setting the recommendation list length as $L=20$, the related results of different methods are reported in table I. We can see that all these metrics are improved by the MD with NM method.

\begin{figure}
  \center
  \includegraphics[width=\columnwidth]{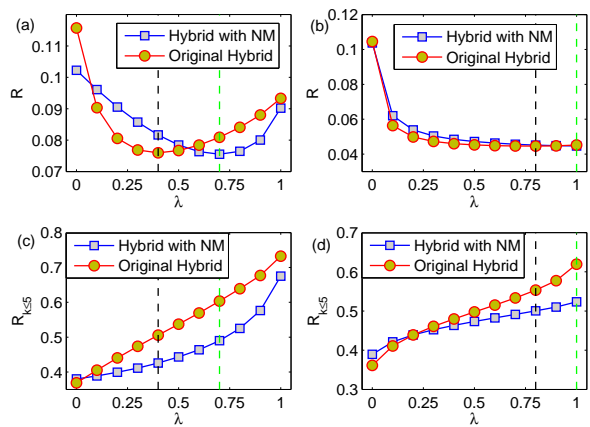}
\caption{(Color online) The overall ranking score $R$ and the local ranking score $R_{k\leq5}$ of the Hybrid with NM method under different $\lambda$ when $Q=0.1$, $\delta=0.1$ in (a), (c) Movielens and $Q=0.1$, $\delta=0.1$ in (b), (d)  Netflix. The black and green vertical dash lines are the optimal $\lambda^*$ for the original Hybrid method and the Hybrid with NM method, respectively.}\label{fig5}
\end{figure}

We then investigate the effect of NM method on the Hybrid algorithm. We first set $Q=0.1$ and $\delta=0.1$ in both data sets, the results are shown in Fig. 5. As we can see, the minimum value of the hybrid method stays almost unchanged after adding the virtual links. However, the $R_{k\leq5}$ under the optimal $\lambda^*$ in the Hybrid with NM method is lower than that in the original hybrid method. The result indicates that the NM method can further solve the item cold-start problem in the Hybrid method without harming at all the overall recommendation accuracy. In our simulation, we try also large $Q$ and $\delta$, the results show that the optimal $R$ for the Hybrid algorithm becomes larger. Therefore, the NM method with relatively small $Q$ and $\delta$ benefits for the Hybrid algorithm. Moreover, we find that the NM method can improve the recommendation diversity of the Hybrid algorithm. The detail value of each metric can be seen in table II.

In the original Hybrid algorithm, the parameter $\lambda$ can adjust amount of score that the Hybrid algorithm gives to the small degree items. When $\lambda$ is close to $0$, the HC algorithm has more weight in the Hybrid algorithm so that the small degree items will obtain more resource from the 3-step diffusion process. However, if a user cannot reach a certain item from the diffusion, the final resource of the item will be always $0$ when $\lambda$ is changed. In NM method, the connectivity of the small degree items can be effectively increased, so that some potential users who are interested in a small degree item might be able to reach such item in the 3-step diffusion. Therefore, NM method can further improve the recommendation accuracy of small degree items in the Hybrid algorithm.

\section{Conclusion}
Information abundant is a serious problem nowadays for online users. In order to filter irrelevant information, many recommendation algorithms have been proposed. In this field, one of the biggest challenges is the item cold-start problem, i.e. the new items have too little historical record to be correctly recommended. So far, all the methods dedicated to solve this problem focused on modifying the existing methods by introducing some parameters. In this letter, we try to solve the problem by directly adding some virtual connection to the bipartite networks so that the niche items have enough information for the recommendation algorithms. Interestingly, besides improving the recommendation accuracy (especially for small degree items), our method can enhance the recommendation diversity compared to the well-known hybrid method of mass diffusion and heat conduction algorithms. In recommender system, note that there are also user cold-start problem and the more difficult cold-start problem for both users and items. These problems have to be addressed in other ways.

In practice, it is actually not necessary to add too many virtual links to the networks to solve the item cold-start problem. Generally, adding $10\%$ links will be sufficient to further enhance the accuracy for those not so popular items in hybrid algorithm. Therefore, our method can be easily applied to real online commercial systems without increasing too much the computational complexity of the recommendation process. Finally, whether the current NM method is the optimal one for each recommendation algorithm is still unknown. For instance, some special algorithms such as the heat conduction algorithm, which mainly recommends niche items, might require for a different virtual link adding strategy. Related problems ask for further investigation in the future.

\acknowledgments
We would like to thank Prof. Yi-Cheng Zhang for helpful suggestions. This work is partially supported by the Foundation of Jiangxi Provincial Department of Education (GJJ. 10696) and National Natural Science Foundation of China under Grant Nos. 71261009.

\end{document}